\documentclass[prd,preprint,superscriptaddress]{revtex4}
\usepackage{revsymb}
\usepackage{amsmath}
\newcommand{\be}{\begin{equation}}
\newcommand{\ee}{\end{equation}}
\newcommand{\ben}{\begin{eqnarray}}
\newcommand{\een}{\end{eqnarray}}

\begin{document}
\title{Dark Energy: A Unifying View}
\date{\today}
\author{Winfried Zimdahl\footnote{Electronic address:
zimdahl@thp.uni-koeln.de}} \affiliation{Departamento de
F\'{\i}sica, Universidade Federal do Esp\'{\i}rito Santo,
CEP29060-900 Vit\'oria, Esp\'{\i}rito Santo, Brazil }
\ \\
\begin{abstract}
Different models of the cosmic substratum which pretend to
describe the present stage of accelerated expansion of the
Universe like the $\Lambda$CDM model or a Chaplygin gas, can be
seen as special realizations of a holographic dark energy
cosmology if the option of an interaction between pressurless dark
matter and dark energy is taken seriously. The corresponding
interaction strength parameter plays the role of a cosmological
constant. Differences occur at the perturbative level. In
particular, the pressure perturbations are intrinsically
non-adiabatic.
\end{abstract}

\maketitle

\section{Introduction}
Since the results of the luminosity distance - redshift
observations of supernovae of type Ia suggested an interpretation
according to which our Universe entered a stage of accelerated
expansion \cite{SNIa}, a host of theoretical concepts has been
developed to account for this phenomenon (for a review see, e.g.,
\cite{rev}). Within Einstein's theory a so far unknown ingredient
with negative pressure is required, which is called dark energy.
By now, many of the dark energy models that were worked out have
been tested against observational data of different kind
\cite{WMAP3,SNIa2,lss}. Different priors and different
parametrizations were used to provide limits on the parameters of
the models under consideration. Still favored is the $\Lambda$CDM
model, but it is also clear that the matter is not solved and that
there are other contenders. In this situation, lacking a
fundamental understanding, one might wish to have a robust
phenomenological framework which allows for a unified description
of (at least a large part of) the currently favored approaches. In
the present essay we demonstrate that holographic dark energy can
provide the basis for such a unifying view. We also point out
unexpected links to cosmological gas dynamics. Finally, we show
that this approach naturally implies the existence of
non-adiabatic pressure perturbations.

\section{The effective equation of state}
Assume the present cosmic substratum to be described by dark
matter and dark energy as the two dynamically relevant components.
In the homogeneous and isotropic, spatially flat Universe
Einstein's equations reduce to
\begin{equation}
3 H^{2} = 8\, \pi G \rho \ ,  \qquad \frac{\dot{H}}{H^{2}} =  -
\frac{3}{2} \left(1 + \frac{p}{\rho}\right)\, , \label{efe}
\end{equation}
where $\rho = \rho_M + \rho_X$ is the total energy density. Here,
$\rho_{M}$ and $\rho_{X}$ are the energy densities of pressureless
dark matter and dark energy, respectively. The pressure of the $X$
component coincides with the total pressure, $p = p_{X}$ and $H$
is the Hubble expansion rate. Solving the last equation in
(\ref{efe}) for $\frac{p}{\rho}$ results in
\begin{equation}\label{wq}
\frac{p}{\rho} = \frac{1}{3} \left(2q - 1\right)\ ,
\end{equation}
where $q =- 1-\frac{\dot H}{H^2}$ is the deceleration parameter.
The matter energy density behaves as (see, e.g., \cite{Das})
\begin{equation}
\rho_{M} =
\rho_{M0}\left(\frac{a_{0}}{a}\right)^{3}\,\frac{f}{f_{0}} \ .
\label{}
\end{equation}
(A subscript 0 denotes the value at the present time.) Here we
have admitted the possibility that the conventional decay of the
matter energy density $\propto a^{-3}$ is modified by an
interaction in the dark sector. Because the total energy has to be
conserved, the density $\rho_{X}$ of the dark energy component
then changes according to
\begin{equation}
\dot{\rho}_{X} = - 3H \left(1 + w^{eff}\right)\rho_{X}\ , \label{}
\end{equation}
where
\begin{equation}
w^{eff} = w + \frac{\dot{f}}{3 H f}\, r\  \label{weff}
\end{equation}
is the effective equation of state parameter while  $w$ is the
corresponding ``bare" parameter and  $r \equiv
\frac{\rho_{M}}{\rho_{X}}$  is the ratio of the energy densities.
In case the energy density ratio is constant, we have
\begin{equation}
r = \mathrm{const} \quad \Leftrightarrow\quad w^{eff} = -
\frac{\dot{f}}{3 H f}\quad \Rightarrow\quad w = \left(1 + r
\right)w^{eff}\, .\ \label{rconst}
\end{equation}
Under this condition the \textit{total} equation of state of the
cosmic medium is
\begin{equation}
\frac{p}{\rho} = w^{eff}\ .\label{pweff}
\end{equation}
It coincides with the effective equations of state of the
components. Apparently, this is a very special situation.
Therefore it may come as a surprise that on this basis many of the
``standard" dark energy models such as the $\Lambda$CDM model or
the Chaplygin gas can be recovered just by different choices of
the interaction. The important aspect here is the following. Via
Friedmann's equation a constant ratio $r$ implies the dependence
$\rho_{X} \propto H^{2}$. While this appears to be an almost
trivial consequence of the relations used so far, the behavior
$\rho_{X} \propto H^{2}$ itself is anything but trivial. It is
exactly this dependence which is found in the context of
holographic dark energy models. The central point of the
holographic dark energy concept is a field theory based relation
between an ultraviolet cutoff and an infrared cutoff \cite{cohen}.
This relation has the attractive feature that, by identifying the
infrared cutoff length with the present Hubble scale, the
corresponding ultraviolet cutoff energy density turns out to be of
the order of the observed value of the cosmological constant
parameter. Just this feature is encoded in the dependence
$\rho_{X} \propto H^{2}$ \cite{cohen}. Despite of this remarkable
property the Hubble scale cutoff has fallen out of favor since for
$f = $ constant it is not consistent with an accelerated expansion
of the Universe. This apparent shortcoming can be remedied and, in
a sense to be pointed out later on, even made an advantage, if the
possibility of an interaction between holographic dark energy and
dark matter is not ignored. The relevance of a coupling between
both components is easily seen. Combining the relations
(\ref{wq}), (\ref{rconst}) and (\ref{pweff}) we obtain
\begin{equation}
q = \frac{1}{2}\left(1 - \frac{\dot{f}}{H f}\right)\ .\label{qf}
\end{equation}
The sign of $q$ crucially depends on the ratio $\frac{\dot{f}}{H
f}$. For $\frac{\dot{f}}{f} < H$ we have $q > 0$, i.e.,
decelerated expansion. For $\frac{\dot{f}}{f} > H$ we have $q < 0$
and accelerated expansion. If, in particular, $f$ is such that the
rate $\frac{\dot{f}}{f}$  changes  from $\frac{\dot{f}}{f} < H$ to
$\frac{\dot{f}}{f}
> H$, this corresponds to a transition from decelerated to
accelerated expansion under the condition of a constant energy
density ratio $r$ (cf. \cite{WD}). This transition is a pure
interaction phenomenon.

\section{Background dynamics}

\subsection{The interaction parameter}

To advance our discussion, information about the rate
$\frac{\dot{f}}{f}$ is required. Since we know neither the nature
of dark matter nor the nature of dark energy, a microphysical
interaction model is not available either. However, one may argue
that under the conditions of spatial homogeneity and isotropy the
only dynamical scale is $H^{-1}$. For the rate $\frac{\dot{f}}{f}$
to be cosmologically relevant it should vary at this scale. It
seems therefore natural to assume a dependence of the crucial
parameter $\frac{\dot{f}}{3 H f}$ in terms of $H^{-1}$. We choose
\begin{equation}
\frac{\dot{f}}{3 H f} = \mu \left(\frac{H}{H_{0}}\right)^{-n}
\qquad \Rightarrow\qquad \dot{\rho} + 3 H\left(1 -
\mu\left(\frac{H}{H_{0}}\right)^{-n}\right)\rho = 0\ . \label{ans}
\end{equation}
The quantity $\mu$ is an interaction constant. Different
interaction rates are characterized by different values of $n$. A
growth of the parameter $\frac{\dot{f}}{H f}$ is obtained for $n
> 0$.
In the spatially flat background the ansatz (\ref{ans})
corresponds to an equation of state parameter
\begin{equation}
\frac{p}{\rho} = - \mu \left(\frac{\rho}{\rho_{0}}\right)^{-n/2}\
. \label{eoszeroo}
\end{equation}
At the present time we have $p_{0}/\rho_{0} = - \mu$, i.e., the
present equation of state parameter is a direct measure of the
interaction parameter $\mu$. Solving the equation for $\rho$ in
(\ref{ans}) we find for the background energy density
\begin{equation}
\rho = \rho_{0}\left[\mu + \left(1 -
\mu\right)\left(\frac{a_{0}}{a}\right)^{3n/2}\right]^{2/n} \ .
\label{rhosolve}
\end{equation}
It has the structure of the energy density of a generalized
Chaplygin gas \cite{Bento}.  In the limit $a \ll a_{0}$ it
reproduces a matter dominated universe with $\rho \propto a^{-3}$,
while in the opposite limit the energy density is similar to that
of a cosmological constant. At first sight this behavior of the
energy density might be unexpected since it was derived under the
condition of a constant ratio $r$ of the energy densities of both
components. However, it is a specific feature of our equation of
state parameter, that the dark energy itself behaves as matter at
high redshifts ($a \ll a_{0}$). At high redshifts we have
$\frac{\dot{f}}{f} \ll H$, i.e.,  the interaction is negligible
(for $n>1$) and we recover a de Sitter universe. It was this
property that apparently ruled out a (non-interacting) holographic
dark energy model with an infrared cutoff set by the Hubble scale
\cite{Hsu,mli1}. Here, this unwanted (in the non-interacting
model) feature is advantageous since it naturally provides us with
an early matter dominated phase during which structure formation
can occur.

For $n = 2$  we recover the $\Lambda$CDM model while for $n = 4$
the expression (\ref{rhosolve}) describes the energy density of a
``true" Chaplygin gas. The cosmological constant term is
determined by the interaction strength parameter $\mu$ of our
approach. This is consistent with the circumstance that for the
Chaplygin gas the parameter which corresponds to $\mu$ represents
the special case of a constant potential term in tachyon field
theories \cite{Frolov}. It is also connected with the interaction
strength of d-branes \cite{Jackiw}. This indicates that there is
support from fundamental field theory for an interaction of the
type introduced through the ansatz (\ref{ans}).

\subsection{Cosmic force}

Another line of understanding the role of the choice (\ref{ans})
emerges if the cosmic medium is studied within a gas dynamical
approach. This provides a suggestion for the origin of
$\frac{\dot{f}}{f} \neq 0$ within kinetic theory.  In this picture
the present phase of accelerated expansion of the Universe is the
result of a cosmic force exerted on the particles of the cosmic
gas \cite{antif}. This force makes the constituents of the cosmic
medium move in a non-geodesic manner while the macroscopic fluid
motion as a whole is geodesic, as required by the cosmological
principle. The equation of motion for the gas particles is
\begin{equation}
\frac{D p^{i}}{d\tau} = mF^{i}\ . \label{}
\end{equation}
Here, $p^{i}$ is the 4-momentum of a particle with mass $m$,
normalized by $p^{i}p_{i} = - m^{2}$ and $\tau$ is its proper
time.  The structure of a 4-force, compatible with the
requirements of the cosmological principle is
\begin{equation}
mF^i = B\left(- E p^i + m^2 u^i\right)\ , \label{Fi}
\end{equation}
with $E \equiv - p^j u_j$ being  the particle energy as measured
by an observer, moving with the macroscopic (geodesic) fluid
4-velocity. The force (\ref{Fi}) contains quantities which
characterize the same fluid both on the microscopic level
(particle momentum, particle energy) and on the macroscopic level
(macroscopic 4-velocity). Hence, it describes a self-interaction
of the medium. The strength of the force is described by the
function $B$. A particle that moves with the geodesic macroscopic
4-velocity is force free,
\begin{equation}
p^i = m u^i\qquad \Rightarrow \qquad F^i = 0
 \ . \label{}
\end{equation}
Any deviation from this motion corresponds to the action of  a
non-vanishing force on the particle. The particles are
characterized by a one-particle distribution function which is
governed by Boltzmann's equation. Assuming the particles to be
non-relativistic, the macroscopic energy balance, obtained from
the second moments of the distribution function, is
\begin{equation}
\dot{\rho} + 3 H\left(1 - \frac{B}{H}\right)\rho = 0
 \ . \label{}
\end{equation}
The correspondence to (\ref{ans}) is obvious,
\begin{equation}
\frac{B}{H}  \Leftrightarrow \mu\left(\frac{H}{H_{0}}\right)^{-n}
 \ . \label{corr}
\end{equation}
With this choice of $\frac{B}{H}$ the energy density
(\ref{rhosolve}) follows from a gas dynamical approach. In other
words, an equation of state parameter $w^{eff}$
(cf.~(\ref{rconst})) can be understood as the result of an
effective self-interacting one-particle force (\ref{Fi}) that
self-consistently acts on the microscopic constituents of the
cosmic substratum. The phenomenologically introduced parameter
$\mu$ is related to the strength of a force on the gas particles.
On the one hand, the relation to the holographic dark energy
concept sheds new light on the cosmic force approach, on the other
hand the dark energy interaction parameter $\mu$ acquires a
counterpart on the level of kinetic theory.

\section{Perturbations}

With $H = \frac{\Theta}{3}$ where $\Theta \equiv u^{i}_{;i}$ is
the fluid expansion scalar, the interaction parameter in
(\ref{ans}) is a covariantly defined quantity. Fluctuations of
this parameter become part of the perturbation dynamics in a
natural way. The quantity $\frac{\dot{p}}{\dot{\rho}}$ which in
standard perfect fluids plays the role of an adiabatic sound speed
is straightforwardly obtained from (\ref{eoszeroo}),
\begin{equation}
\frac{\dot{p}}{\dot{\rho}} = \left(1 -
\frac{n}{2}\right)\frac{p}{\rho}
 \ . \label{}
\end{equation}
However, it is \textit{not} this quantity that relates the
pressure perturbations to the energy density perturbations in our
approach. The pressure perturbations in linear order are (cf.
(\ref{rconst}), (\ref{pweff}) and (\ref{ans}))
\begin{equation}
\hat{p} = p \left(\frac{\hat{\rho}}{\rho} - n
\frac{\hat{H}}{H}\right)
 \ . \label{pr}
\end{equation}
Quantities without a hat refer to the homogeneous and isotropic
background in the following. The perturbation $\hat{H}$ of $H$ is
defined via the expansion scalar $\Theta$ as $\hat{H} =
\frac{\hat{\Theta}}{3}$. The pressure perturbations (\ref{pr}) are
not simply proportional to the energy density perturbations. This
is a consequence of the circumstance that there exists an equation
of state $p = p(\rho)$ only in the background (cf.
Eq.~(\ref{eoszeroo})) but not for deviations from homogeneity and
isotropy. The fluctuations of the interaction parameter make the
perturbations non-adiabatic. The deviation from adiabatic behavior
is most conveniently described by
\begin{equation}
\hat{p} - \frac{\dot{p}}{\dot{\rho}} \hat{\rho} = \frac{n}{2} p
\left(\frac{\hat{\rho}}{\rho} - 2 \frac{\hat{H}}{H}\right)
 \ . \label{hatp}
\end{equation}
There are no non-adiabatic contributions only for $n = 0$. The
combination $\frac{\hat{\rho}}{\rho} - 2\frac{\hat{H}}{H}$ on the
right hand side of (\ref{hatp}) is proportional to the perturbed 3
curvature scalar $R^{(3)}$ of the surfaces orthogonal to $u^{i}$.
In first order we have in the present case
\begin{equation}
\hat{R}^{(3)} = 6\,H^{2}\left(\frac{\hat{\rho}}{\rho} - 2
\frac{\hat{H}}{H}\right)
 \, .  \label{}
\end{equation}
Thus, the non-adiabatic pressure perturbations of our approach
have a direct geometrical meaning. In terms of (gauge invariant)
perturbation quantities on comoving hypersurfaces, a combination
of the energy and momentum balances of the fluid can be used to
eliminate the perturbations of the Hubble parameter. The result is
\begin{equation}
\hat{p} = \frac{p}{\rho} \left[\left(1 +
\frac{n}{\gamma}\right)\hat{\rho}
 + \frac{n}{3\gamma H}\dot{\hat{\rho}}\right] \ ,\qquad
\qquad\left(\gamma = 1 + \frac{p}{\rho}\right)
 \ . \label{hatpfin}
\end{equation}
The  remarkable point here is  that the pressure perturbations are
not just proportional to the energy density perturbations
$\hat{\rho}$ as in the adiabatic case. There is an additional
dependence on the time derivative $\dot{\hat{\rho}}$ of the energy
density perturbations. The relation between $\hat{p}$ and
$\hat{\rho}$ is no longer simply algebraic, equivalent to a
(given) sound speed parameter as a factor relating the two. The
relation between them becomes part of the dynamics. In a sense,
$\hat{p}$ is no longer a ``local" function of $\hat{\rho}$ but it
is a function of the derivative $\dot{\hat{\rho}}$ as well:
$\hat{p} = \hat{p}(\hat{\rho}, \dot{\hat{\rho}})$. It is only for
the background pressure that the familiar dependence $p = p(\rho)$
is retained. Formula (\ref{hatpfin}) is a direct consequence of
the structure (\ref{ans}) for the interaction parameter
$\frac{\dot{f}}{H f}$. While this interaction reproduces known
dark energy models in the homogeneous and isotropic background,
albeit in a non-standard unifying context, there are differences
on the perturbative level which opens the possibility to test the
scheme presented here.

\section{Summary}

Pressureless dark matter in interaction with holographic dark
energy with an infrared Hubble scale cutoff is more than just
another model to describe an accelerated expansion of the
Universe. It sets the stage for a unifying view on a whole class
of models, among them the $\Lambda$CDM model and the Chaplygin gas
model, which follow as subcases for different interaction rates.
The interaction can be interpreted in terms of a 4-force on the
constituents of the cosmic gas. The unifying view on the
homogeneous and isotropic background  is accompanied by a
non-adiabatic perturbation dynamics which can be seen as the
consequence of a fluctuating interaction rate. The relation
between pressure perturbations and energy density perturbations
becomes part of the dynamics and is no longer given by a simple
sound speed parameter.

\end{document}